\documentclass[pra,twocolumn,floatfix,superscriptaddress]{revtex4-1}
\usepackage{dcolumn,amsmath}
\usepackage{graphicx}
\usepackage{bm}
\usepackage{amssymb}
\usepackage{multirow}

\usepackage{amsfonts}
\usepackage{amsmath}
\usepackage{cancel}
\usepackage{xcolor}
\newcommand{\ecm}{\ensuremath{e {\cdotp} {\rm cm}}}

\newcommand{\de}{d_\mathrm{e}}

\begin{document}
\title{Hyperfine structure and $K$-doubling in RaOCH$_3$ molecule}
\author {Alexander Petrov}\email{petrov\_an@pnpi.nrcki.ru}
\affiliation{Petersburg Nuclear Physics Institute named by B.P. Konstantinov of National Research Centre
"Kurchatov Institute", Gatchina, 1, mkr. Orlova roshcha, 188300, Russia}
\affiliation{St. Petersburg State University, St. Petersburg, 7/9 Universitetskaya nab., 199034, Russia} 

\date{Received: date / Revised version: date}
%
\begin{abstract}{
 RaOCH$_3$ molecule is one of the most sensitive systems for the electron electric dipole moment ($e$EDM) searches. Its hyperfine and $K-$doubling structure in the external electric field is of key importance for preparing and interpreting the experiment. We propose the method for calculation the symmetric top molecules and applied it for RaOCH$_3$. Hyperfine structure, $K-$doubling and sensitivity of the molecule to $e$EDM in the external electric field were calculated. } 
\end{abstract}
\maketitle
\section{Introduction}

Measuring the electron electric dipole moment ($e$EDM) is a most promising test for existence of physics beyond the Standard model.
  Its value 
  measured at the level of the current experimental sensitivity would be a clear signature of the physics beyond the Standard model (SM) 
\cite{KL95,GFreview,Titov:06amin, Feng:2013, Safronova:18}.
Recently the JILA group has obtained a new constraint on the electron electric dipole moment ($e$EDM), $|\de|<4.1\times 10^{-30}$ \ecm\ (90\% confidence)
  \cite{newlimit1},
using the $^{180}$Hf$^{19}$F$^+$ ions trapped by the rotating electric field. 
The measurement was performed on the ground rotational sublevel of the metastable first excited electronic $^3\Delta_1$ state.
It further improves the latest ACME collaboration result obtained in 2018, $|d_e| \lesssim 1.1\cdot 10^{-29}\ e\cdot\textrm{cm}$ \cite{ACME:18}, by a factor of 2.4 and the first result $|\de|\lesssim 1.3\times 10^{-28}$ on the $^{180}$Hf$^{19}$F$^+$ ions  \cite{Cornell:2017} by a factor of about 32.

 The great progress in $e$EDM search on HfF$^+$  and ThO is closely related to $\Omega-$doubling structure (close levels of opposite parities) of these molecules, which helps to suppress many systematic effects \cite{DeMille2001}.

It is clear now that in experiments with laser-cooled molecules the statistical sensitivity can be strongly enhanced as a result of the increasing of the interrogation time \cite{Isaev:16, Isaev_2017}. Unfortunately, the complicated electronic structure of $\Omega-$doublets prevents laser cooling of corresponding diatomic molecules.

However, as first pointed out in Ref. \cite{Kozyryev:17}, the triatomic molecules with linear equilibrium configuration (RaOH, YbOH, etc.) and symmetric top molecules (RaOCH$_3$, YbOCH$_3$, etc.) possess $l-$doubling and $K-$doubling structures respectively, which play the same role as $\Omega-$doubling one in diatomics for suppressing the systematic effects, but being not related to electronic structure do not prevent the laser cooling.
$l$-doubling structure in triatomics is related to two degenerate bending (distorting linear configuration) vibrational modes in orthogonal planes in the {\it excited} $v=1$ ($v$ is a vibrational quantum number) vibrational mode. $K-$doubling structure in symmetric top molecules is related to two different projections ($K$) of the rotational momentum on the symmetry axis.

Great progress has been achieved recently in both theoretical and experimental studies of triatomics \cite{Kozyryev:17, Steimle:2019, Augenbraun_2020, zakharova21b, zakharova21, PhysRevA.103.022813, Maison:22, Kurchavov:22, Petrov:2022, zakharova22, kurchavov23,  Persinger:2023, Jadbabaie_2023, anderegg2023quantum, Petrov:24a,Petrov:24b}. Meantime the sensitivity to $e$EDM and other $\mathcal{T,P}$-odd ($\mathcal{P}$ -- spatial parity and $\mathcal{T}$ time reversion) effects in symmetric top molecules is less studied. In Refs. \cite{Zhang:21, zakharova22, Gaul:24} the $\mathcal{T,P}$-odd electronic structure parameters $W_d$ and $W_s$ for BaOCH$_3$, YbOCH$_3$, RaOCH$_3$, RaOCH$_3^+$, BaCH$_3$ and YbCH$_3$ were calculated. In Ref. \cite{Augenbraun:21} the laser spectroscopic study of YbOCH$_3$ was performed and rapid optical cycling was found for laser cooling. In Ref. \cite{Fan:25} the mass spectrometry technique to identify trapped ions on example of RaOCH$_3^+$ were presented. In Ref. \cite{Yu:2021}
the detailed theoretical study of the closed shell RaOCH$_3^+$ for precision measurements of Schiff moment was performed. The  hyperfine and $K-$doubling structures and polarizability in the external electric field were calculated. To the best of our knowledge, hyperfine and $K-$doubling structures and sensitivities to $\mathcal{T,P}$-odd effects in the external electric field were not yet studied for the open shell symmetric top molecules. The main goal of the present Letter is to consider these effects on the example of RaOCH$_3$ molecule.

\section{Theory}
\subsection{Hyperfine and $K-$doubling structure}
The rotational levels of the ground electronic-vibrational state of RaOCH$_3$ are well described by the Hund's case $b$ coupling scheme. Electron spin ${S=1/2}$ for a good approximation is an integral of motion. Its interaction (spin-rotation) with the rotational momentum $\hat{\bf N}$ gives rise to the splitting between the energy levels with total electronic-rotational $J = N \pm 1/2$ momenta. 
We consider the RaOCH$_3$ molecule with spinless isotopes of Ra, O and C, whereas the hydrogens have a non-zero nuclear spins $i{=}1/2$. The total spin of the hydrogen nuclei is $I=1/2$ (two states) and $I=3/2$ (one state), which gives rise to the hyperfine energy splitting between the levels with the total (electronic-rotational-nuclear spin) angular momentum $\hat{\bf F}=\hat{\bf J}+\hat{\bf I}$. 
The focus of the $e$EDM experiment is the $N=1$ rotational level, which is the lowest one having $K-$doubling structure. Taking into account that for $N=1$ the quantum number $K$ can have values $-1,0,1$, one can calculate that $N=1$ level has 144 different states. 
They are three $F=3$, twelve $F=2$, eighteen $F=1$ and nine $F=0$. 
However, the discussion above does not take into account the Pauli principle for identical hydrogen nuclei. 
The allowed by Pauli principle states, their
irreducible representations of the group $C_{3v}$ (the point group of RaOCH$_3$) for the rotational wavefunctions (electronic and vibrational ones are assumed be totally symmetric), nuclear statistical factors and parities of the states are given in Table \ref{psym} \cite{Land3}. It should be noted that $C_{3v}$ group contains reflection in the planes passing through the symmetry axis, and it is pointless to consider transformation of the rotational wavefunction under reflection. Therefore, reflection in the plane should be understood as a rotation on $\pi$ around axis perpendicular to the reflection plane followed by inversion \cite{Land3}. Inversion changes sign for all coordinates of electrons and nuclei and produces another configuration of the RaOCH$_3$ molecule which can not be obtained (as oppose to diatomics, for example) by any rotation of the original configuration. Definite parity state is a combinations of these two configurations. Rotational wavefunction of $A_2$ symmetry corresponds to $I=3/2$ ($I=3/2$ transforms according to $A_1$) and rotational wavefunction of $E$ symmetry corresponds to $I=1/2$ (two $I=1/2$ states transforms according to $E$).
The total wavefunction including nuclear spin variables transforms according to $A_2$ representation which ensures the implementation of the Pauli principle. According to Table \ref{psym}, for $N=1$ we have one $F=3$, two $F=2$, two $F=1$, one $F=0$ for $I=3/2, K=0$ and two $F=2$, four $F=1$, two $F=0$ for $I=1/2, |K|=1$. So, after taking into account the Pauli principle the total number of states is reduced from 144 to 48. It is clear from Table \ref{psym} that states with $I=1/2$ ($E$ symmetry) are just $K-$doublets of opposite parities. Due to extreme closeness of these levels they will be polarized by a very small electric field (see below), and are of interest for $e$EDM search experiment. $M_F$= 0 ($M_F$ is projection of the total momentum on laboratory axis) are insencitive to $e$EDM. So, the focus of the current Letter will be on two (one $K-$doublet) $M_F=2$ levels and six (three $K-$doublets) $M_F=1$ levels.

\begin{table}
\caption{Allowed by Pauli principle rotational states, their symmetry and nuclear statistical factors. See text for details}
\begin{tabular}{lcc}
\hline
&  + & - \\
$|K|$ {not a multiple of} 3 & $2E$ & $2E$ \\
$|K|$ {  a multiple of} 3    & $4A_2$ & $4A_2$ \\
K=0, N even                   &  --   &  $4A_2$ \\
K=0, N odd                    & $4A_2$ &   -- \\
\hline
\end{tabular}
\label{psym}
\end{table}

\subsection{Method} 
\begin{figure}[h]
\centering
  \includegraphics[width=0.25\textwidth]{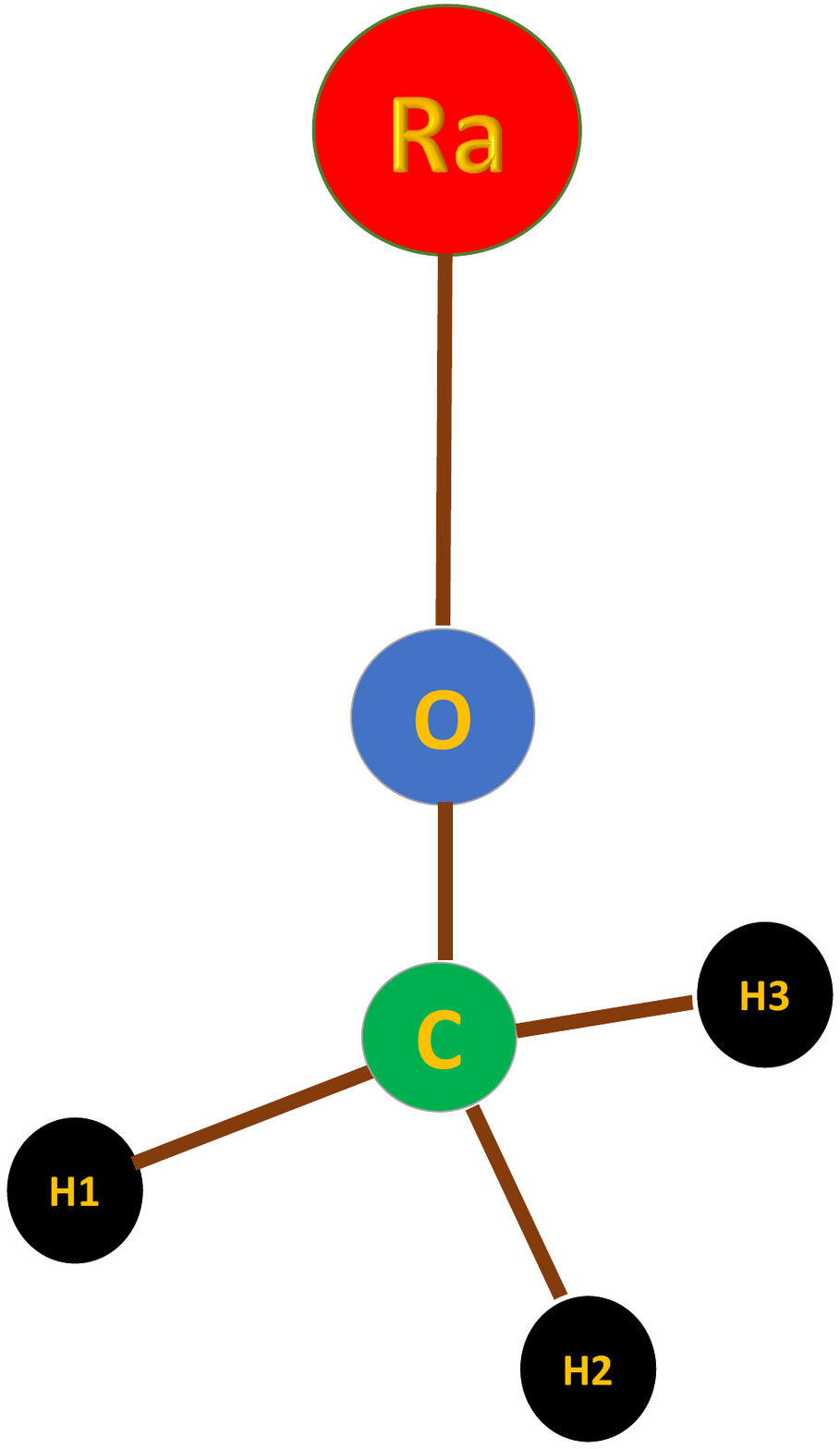}
  \caption{Symmetric top RaOCH$_3$ molecule}
  \label{symtop}
\end{figure}

We present total Hamiltonian in molecular reference frame as
\begin{equation}
{\rm \bf\hat{H}} = {\rm \bf\hat{H}}_{\rm mol} + {\rm \bf\hat{H}}_{\rm hfs} + {\rm \bf\hat{H}}_{\rm ext},
\label{Hamtot}
\end{equation} 
where
\begin{equation}
\hat{\rm H}_{\rm mol}= B(\hat{\bf J} -\hat{\bf J}^{e} )^2+ \left(A-B\right)(\hat{J}_z -\hat{ J}^{e}_z)^2
\label{Hmolf}
\end{equation}
is the molecular Hamiltonian, $B= 0.0631$ cm$^{-1}$ and $A=4.83$ cm$^{-1}$ are rotational constants corresponding to the equilibrium geometry obtained in Ref. \cite{zakharova22},
$\hat{\bf J}^{e}$  is the electronic momentum.  So, in the current work we consider RaOCH$_3$ as a rigid symmetric top depicted in Fig \ref{symtop}.

 \begin{equation}
 {\rm \bf\hat{H}}_{\rm hfs} = { g}_{\rm H} \sum_{i=1}^3 {\bf \rm I^{i}} \cdot \sum_a\left(\frac{\bm{\alpha}_a\times \bm{r}_{ia}}{r_{ia}^3}\right)
\end{equation}
is the hyperfine interaction between electrons and the hydrogen nuclei,
${ g}_{\rm H}$ is the
 g-factor of the hydrogen nucleus, $\bm{\alpha}_a$
 are the Dirac matrices for the $a$-th electron, $\bm{r}_{ia}$ is its
 radius-vector in the coordinate system centered on the H nuclei,
index $a$ enumerates electrons of RaOCH$_3$.

\begin{equation}
 {\rm \bf\hat{H}}_{\rm ext} =   -{ {\bf D}} \cdot {\bf E}
\end{equation}
describes the interaction of the molecule with the external electric field ${\bf E}$, and
{\bf D} is the dipole moment operator.

Wavefunctions and energies were obtained by numerical diagonalization of the Hamiltonian (\ref{Hamtot})
over the basis set of the electronic-rotational-vibrational-nuclear spins wavefunctions.
We start from the basis set
\begin{equation}
 \Psi^e_{\Omega }\Theta^{J}_{M_J,\omega}(\alpha,\beta,\gamma)U^{1}_{M^{1}_I}U^{2}_{M^{2}_I}U^{3}_{M^{3}_I}.
\label{basis}
\end{equation}
Here 
 $\Theta^{J}_{M_J,\omega}(\alpha,\beta, \gamma)=\sqrt{(2J+1)/{8\pi^2}}D^{J}_{M_J,\omega}(\alpha,\beta,\gamma)$ is the rotational wavefunction, $\alpha,\beta, \gamma$ are Euler angles defining orientation of the molecular frame $xyz$ relative to the laboratory one XYZ.
 $z$ axis corresponds to the symmetry axis of RaOCH$_3$, $x$ axis oriented in such a way that the first hydrogen atom lies in the $xz$ plane.
 $U^{i}_{M^{i}_I}$ are the hydrogen nuclear spin wavefunctions, $M_J$ is the projection of the molecular (electronic-rotational) angular momentum $\hat{\bf J}$ on the $Z$ axis, 
 $\omega$ is the projection of the same momentum on the $z$ axis,
 $M^{i}_I$  are the projections of the hydrogen nuclear angular 
momenta  on the $Z$ axis,  
$\Psi^e_{\Omega}$ is the electronic wavefunction, $\Omega \approx \pm 1/2$ is the projection of the 
electronic angular momentum on $z$ axis. Then, using the approximate relation $K=\omega-\Omega$, we construct proper antisymmetric basis functions for $N=0,1,2$ according to Table \ref{psym}.

To the best of our knowledge, calculation of all required diagonal and off-diagonal {\it electronic} matrix elements, using the methods accounting for electron correlation effects, is not currently available in public quantum-chemical codes and should be a goal for the further development. In the current Letter we used the Dirac 19 software package \cite{DIRAC19} to calculate molecular orbitals using the self-consistent field (SCF) method with
a ten-electron generalized relativistic effective core potential (GRECP) \cite{titov1999generalized,mosyagin2010shape,mosyagin2016generalized} including spin-orbit interaction  with associated basis set \cite{QCPNPI:Basis}. The cc-pVTZ basis sets were employed for O, C and H atoms.
Required matrix elements for $\hat{J}^{e}_z$,  $\hat{J}^{e}_+ = \hat{J}^{e}_x + i\hat{J}^{e}_y$,
\begin{equation}
\nonumber
 \hat{A}_{\parallel}^{(i)} = \frac{{ g}_{\rm H}}{\Omega} \sum_a\left(\frac{\bm{\alpha}_a\times \bm{r}_{ia}}{r_{ia}^3}\right)_z,
\end{equation}
\begin{equation}
\nonumber
 \hat{A}_{\perp}^{(i)} = { g}_{\rm H} \sum_a \left[ \left(\frac{\bm{\alpha}_a\times \bm{r}_{ia}}{r_{ia}^3}\right)_x + i \left(\frac{\bm{\alpha}_a\times \bm{r}_{ia}}{r_{ia}^3}\right)_y \right]
\end{equation}
are given below as matrices where the first basis function is $\Psi^e_{\Omega=-1/2}$ and the second one is
$\Psi^e_{\Omega=1/2}$. 

\subsection{Interaction with $e$EDM}
The $e$EDM interaction is described by the Hamiltonian
\begin{eqnarray}
  \hat{\bf H}_{\rm d}=2d_e
  \sum_a\left(\begin{array}{cc}
  0 & 0 \\
  0 & \bm{\sigma}_{ a} \bm{E}_{in} \\
  \end{array}\right)\ ,
 \label{Hd}
\end{eqnarray}
where $d_e$ is the value of electron electric dipole moment, $\bm{E}_{in}$ is the inner molecular electric field (do not mix with external electric field $\bm{E}$), and $\bm{\sigma}$ are the Pauli matrices. 
In the absence of external electric field each state has definite parity and average value of Hamiltonian (\ref{Hd}) is zero, so molecule is insensitive to the $e$EDM. $K-$doublets levels of opposite parities 
 present equal-mixed combinations of $K=\pm1$ as well as $\Omega=\pm1/2$ and have slightly different energies. External electric field mixes levels of opposite parities and corresponding eigenstates become sensitive to the $e$EDM. 
 Any $e$EDM experiment just searches for the corresponding energy shift  ($e$EDM induced Stark shift),
 which can be calculated as
\begin{equation}
\delta E =\langle\Psi\left| \hat{\bf H}_{\rm d} \right| \Psi \rangle \equiv P E_{\rm eff} d_e,
\label{split}
\end{equation}

\begin{equation}
\label{matrelem}
E_{\rm eff} = W_d/2 = \frac{1}{d_e}
\langle \Psi_{\Omega=1/2}| \hat{\bf H}_{\rm d}|\Psi_{\Omega=1/2}\rangle,
\end{equation}
where $\Psi$ is eigenfunction of Hamiltonian (\ref{Hamtot}), $E_{\rm eff}$ is {\it effective electric field} acting on electron in the molecule,
$P$ is the polarization of the molecule by the external electric field.
 (We note, that $P$, in general, is not equal to the mean value of the projection of unit vector $\hat{z}$ along molecular axis on direction of the external electric field, see below).
To extract $d_e = \delta E / (E_{\rm eff} P) $ from the measured
shift $\delta E $, one needs to know $PE_{\rm eff}$.
 $E_{\rm eff}$ was a subject of previous molecular calculations.
Obtained data $E_{\rm eff} =54$ GV/cm \cite{Zhang:21} and $E_{\rm eff} =48$ GV/cm  \cite{zakharova22} are in good agreement with $E_{\rm eff} =46$ GV/cm estimated in this Letter by the SCF method. $P$ was not previously calculated and is one of the focus of the present Letter. It is clear from Eq. (\ref{split}) that $P$ can be formally calculated as
\begin{equation}
P = \frac{1}{E_{\rm eff} d_e}  \langle\Psi\left| \hat{\bf H}_{\rm d} \right| \Psi \rangle.
\label{Pe}
\end{equation}
The states connected by the time reversal $M_F \rightarrow -M_F$ have the opposite $e$EDM induced Stark shifts (Actually $e$EDM experiment searches for the corresponding energy splitting), which means that $P(M_F) = -P(-M_F)$ for these states. Therefore, we focus on positive values $M_F = 1$ and $M_F = 2$. Note, that for the $K-$doublets we have $P(M_F) \approx -P(M_F)$, see Fig. \ref{EDMshift} below.

\section{Results}
The obtained electronic matrix elements are
%
%
\begin{equation}
\hat{J}^{e}_z = 
\begin{pmatrix}
-0.49986 &  0.0000 \\
0.0000 &  0.49986
\end{pmatrix},
\label{Jz}
\end{equation}
%
%
\begin{equation}
\hat{J}^{e}_+ = 
\begin{pmatrix}
0.0000 &  0.0000 \\
 1.0342 & 0.0000
\end{pmatrix},
\label{Jp}
\end{equation}
%
%
\begin{equation}
\hat{A}_{\parallel}^{(1)} = 
\begin{pmatrix}
0.6034 &  0.1275 \\
0.1275 &  -0.6034
\end{pmatrix} {\rm MHz},
\label{Az1}
\end{equation}
%
%
\begin{equation}
\hat{A}_{\perp}^{(1)} = 
\begin{pmatrix}
-0.0863 & -0.0392 \\
 0.0884 &  0.0863
\end{pmatrix} {\rm MHz},
\label{Ap1}
\end{equation}
%
%
\begin{equation}
\hat{A}_{\parallel}^{(2)} = 
\begin{pmatrix}
0.6034 &   -0.0637 + 0.1104i \\
 -0.0637 -0.1104i &  -0.6034
\end{pmatrix} {\rm MHz},
\label{Az2}
\end{equation}
%
%
\begin{equation}
\hat{A}_{\perp}^{(2)} = 
\begin{pmatrix}
0.0432 -0.0748i & 0.0196 + 0.0339i \\
 0.0884 &  -0.0432 + 0.0748i
\end{pmatrix} {\rm MHz},
\label{Ap2}
\end{equation}
%
%
\begin{equation}
\hat{A}_{\parallel}^{(3)} = 
\begin{pmatrix}
0.6034 &   -0.0637 - 0.1104i \\
 -0.0637 + 0.1104i &  -0.6034
\end{pmatrix} {\rm MHz},
\label{Az3}
\end{equation}
%
%
\begin{equation}
\hat{A}_{\perp}^{(3)} = 
\begin{pmatrix}
0.0432 + 0.0748i & 0.0196 - 0.0339i \\
 0.0884 &  -0.0432 - 0.0748i
\end{pmatrix} {\rm MHz},
\label{Ap3}
\end{equation}
Note that $\hat{J}^{e}_+$ and $\hat{A}_{\perp}^{(i)}$ operators are not hermitian and can have complex diagonal matrix elements. As a test of the calculated matrix elements, we verified that $\hat{A}_{\parallel}^{(i)}$, $\hat{A}_{\perp}^{(i)}$ can be obtained for one $i$ from another by transformation of $\Psi_{\Omega=-1/2}$, $\Psi_{\Omega=1/2}$ spinors and $\hat{A}_{x}, \hat{A}_{y}, \hat{A}_{z}$ vectors,
when rotating around the $z$ axis on $\pm 120$ degrees. As a test of our hyperfine structure calculations, we verified that we obtain identical energy levels when nullifying hyperfine interactions with any two hydrogen nuclei.

Calculated values for energy splitting of $K-$doublets are 13.6 kHz for $N=1,J=3/2,F=1$,  8.2 kHz for  $N=1,J=3/2,F=2$ 
and 26.1 kHz for $N=1,J=1/2,F=1$. We note, that current values should be considered as estimation by the order of magnitude. Since wavefunctions of $K-$doublets have structure  $(|K=+1, 1I=1/2\rangle \pm |K=-1, 2I=1/2\rangle)$, and $1I=1/2$, $2I=1/2$ hydrogen spin wavefunctions are orthogonal, we do not see any interaction except hyperfine one, which can lead to notable $K-$doubling.
Since hyperfine interaction is small, the main effect is due to the hyperfine interaction between  $|K=\pm1\rangle$ states, which due to the $K-$doubling structure gives opposite energy shifts for the components of $K-$doublet. This interaction is nonzero due to matrix elements 
 $ (\hat{A}_{\perp}^{(i)})_{12} =  \langle\Psi^e_{\Omega=-1/2} | \hat{A}_{\perp}^{(i)} |\Psi^e_{\Omega=1/2} \rangle$, which violates symmetry of the linear configuration.

\begin{figure}
\includegraphics[width=0.95\linewidth]{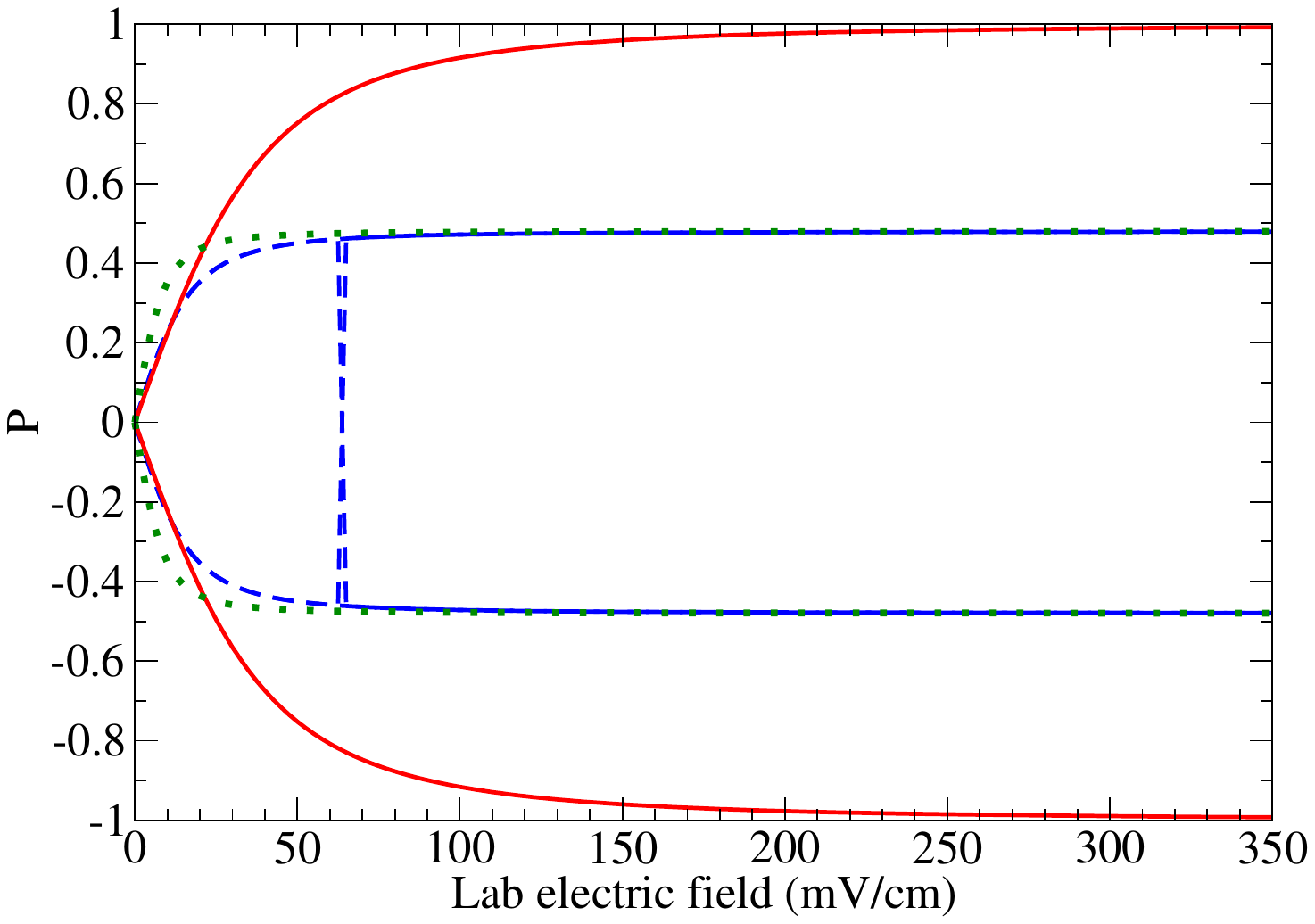}
\caption{\label{EDMshift} 
(Color online) Calculated polarization $P$ (see eq. (\ref{split})) for the $K-$ doublets with dipole moment $D=0.7$ a.u.  Solid (Red) lines: $J=1/2, F=1, M_F=1$, Dashed (blue) lines: $J=3/2, F=1,2, ~M_F=1$. Curves for $F=1$ and $F=2$, except small region of avoiding crossing, are indistinguishable on the figure. Dotted (green) lines $J=3/2, F=2 M_F=2$.}
\end{figure}

 In Fig. \ref{EDMshift} the calculated
polarizations $P$ for six $M_F = 1$ and two $M_F = 2$  hyperfine sublevels 
of the $K-$doublet of $N = 1$ rotational level are given.
Calculations show that most of the levels smoothly approach $P \approx 0.48$ value.
This fact can be qualitatively explained in the Hund's case (b) coupling scheme.
First, recall that $e$EDM is directed along the electron spin. Projection of the later on the laboratory axis, for a good approximation, is integral of motion. Internal effective electric field of RaOCH$_3$ is directed along the molecular axis $z$. Thus, projection of the corresponding unit vector $\hat{z}$ on laboratory axis will give the $P$ value. Using the expression $\langle NMK |\hat{z}| NMK\rangle = MK/N(N+1)$ \cite{Land3} (M is the projection of {\bf N} on laboratory axis), for $N,|M|,|K|=1$ immediately obtain $|P|=0.5$. 

However, this consideration does not take into account spin-rotation interaction and, therefore, does not explain $|P|=1$ value for $N=1,J=1/2,F=1$.
The spin-rotation interaction couples $\hat{\bf N}$ and $\hat{\bf S}$ to the $\hat {\bf J}$ angular momentum. In the Hund's case (c) coupling scheme the corresponding states are
\begin{eqnarray}
\nonumber
| K=1,N=1,S,J=3/2,M_J\rangle = \\
\nonumber
\sqrt{3/4}|\Omega=1/2,K=1,J=3/2,M_J\rangle + \\
 \sqrt{1/4}|\Omega=-1/2,K=1,J=3/2,M_J\rangle, 
 \label{WF32}
\end{eqnarray}
\begin{eqnarray}
\nonumber
| K=1,N=1,S,J=1/2,M_J\rangle = \\
|\Omega=-1/2,K=1,J=1/2,M_J\rangle.
\label{WF12}
\end{eqnarray}
$| \Omega=1/2 \rangle$ and $| \Omega=-1/2 \rangle$ states have opposite contribution to the $e$EDM induced Stark shift, therefore $|P|$ can be calculated as a difference of weights of these states.
Thus, we have $|P|=1$ for $J=1/2$ and $|P|=3/4-1/4=1/2$ for $J=3/2$.

Consideration in the Hund's case (b) coupling scheme corresponds to sufficiently large electric field when $\hat{\bf N}$ and $\hat{\bf S}$ are uncoupled. In in the Hund's case (c) coupling scheme this corresponds to the mixing by external electric field of $J=3/2$ and $J=1/2$ states.
Wavefunctions (\ref{WF32}) and (\ref{WF12}) have definite $K$ quantum number, which imply complete mixing of the $K-$doublet components but ignore mixing of $J=3/2$ and $J=1/2$. In reality components of $K-$doublet and $J=1/2$ and $J=3/2$ mix simultaneously and $|P|$ of $J=1/2$ instead of approaching the unit value goes through maximal value $0<|P|_{\rm max}<1$ and then finaly saturated at $|P|=0.5$ value. The picture is very similar to triatomic molecules and can be seen  for curve number 6 at Fig. 2 in Ref. \cite{Petrov:2022}. It is clear from consideration above that the larger the spin-rotation interaction and the smaller the $K-$doubling the larger will be $|P|_{\rm max}$ and the wider will be the maximum.
For RaOCH$_3$, due to the very small value of $K-$doubling, $|P|_{\rm max}\approx 1$ and maximum is wide.
On Fig \ref{EDMshift} we observe it as a saturation of $|P|$ at unit value for $J=1/2$. For $E=100$ V/cm $P=0.61$ and $P=-0.83$ for $K-$doublet components of $J=1/2$. 

$J=3/2$ is also obtained when spin couples with $N=2$ rotational state. For this state we can write
\begin{eqnarray}
\nonumber
| K=1,N=2,S,J=3/2,M_J\rangle = \\
\nonumber
-\sqrt{1/4}|\Omega=1/2,K=1,J=3/2,M_J\rangle + \\
 \sqrt{3/4}|\Omega=-1/2,K=1,J=3/2,M_J\rangle. 
 \label{WF32b}
\end{eqnarray}
If matrix element $(\hat{J}^{e}_+)_{21} = 1$, then wavefunctions (\ref{WF32}) and (\ref{WF32b}) do not interact. This corresponds to the free electron (zero spin-rotational interaction) and to saturated value $|P|=0.5$ for $J=3/2$. One can verify that if $(\hat{J}^{e}_+)_{21} > 1$ ( $(\hat{J}^{e}_+)_{21} < 1$), then mixing of wavefunctions leads to $|P|<0.5$ ($|P|>0.5$). 
In our case $(\hat{J}^{e}_+)_{21} = 1.0342$ (see Eq. (\ref{Jp})), which explains the value $|P| \approx 0.48 < 0.5$ for $J=3/2$ mentioned above.

\section{Conclusion}
We developed the method for calculation of the hyperfine structure of the symmetric top molecules. Electronic matrix elements required for calculation of the $K-$doubling are determined. $K-$doubling for the first excited rotational level of RaOCH$_3$
were estimated on the levels of  13.6 kHz for $N=1,J=3/2,F=1$,  8.2 kHz for  $N=1,J=3/2,F=2$ and 26.1 kHz for $N=1,J=1/2,F=1$. $\mathcal{T},\mathcal{P}-$odd polarization determining sensitivity to $e$EDM is calculated as function of the external electric field. 

\section{Acknowledgements}
Calculations of RaOCH$_3$ are supported by the Russian Science Foundation grant no. 24-12-00092. Electronic structure calculations were carried out using computing resources of the federal collective usage center Complex for Simulation and Data Processing for Mega-science Facilities at National Research Centre ``Kurchatov Institute'', http://ckp.nrcki.ru/.
%
%

\end{document}